# Mapache: a flexible pipeline to map ancient DNA

Samuel Neuenschwander [1,2†*], Diana I. Cruz Dávalos [1,3†], Lucas Anchieri [1,3], Bárbara Sousa da Mota [1,3], Davide Bozzi [1,3], Simone Rubinacci [1,3], Olivier Delaneau [1,3], Simon Rasmussen [4], and Anna-Sapfo Malaspinas [1,3]

[1] Department of Computational Biology, University of Lausanne, Switzerland
[2] Vital-IT, Swiss Institute of Bioinformatics, Lausanne, Switzerland
[3] Swiss Institute of Bioinformatics, Lausanne, Switzerland
[4] Novo Nordisk Foundation Center for Protein Research, University of Copenhagen, Denmark

[†] Contributed equally.
[*] To whom correspondence should be addressed.

**Abstract**
**Summary:** Mapping ancient DNA to a reference genome is challenging as it involves numerous steps, is time-consuming and has to be repeated within a study to assess the quality of extracts and libraries; as a result, the mapping needs to be automated to handle large amounts of data in a reproducible way.
We present mapache, a flexible, robust, and scalable pipeline to map, quantify and impute ancient and present-day DNA in a reproducible way. Mapache is implemented in the workflow manager Snakemake and is optimized for low-space consumption, allowing to efficiently (re)map large data sets such as reference panels and multiple extracts and libraries.
**Availability**: Mapache is freely available on GitHub (https://github.com/sneuensc/mapache).
**Contact**: samuel.neuenschwander@unil.ch
**Supplementary information:** An extensive manual is provided at https://github.com/sneuensc/mapache/wiki.

**Introduction**

Mapping sequencing reads to a reference genome is a fundamental step in genomic analyses. Compared to modern data, ancient DNA (aDNA) presents a number of challenges as data is sparse, contaminated, and contains post-mortem damage, such as fragmentation and deamination. Furthermore, samples are often re-sequenced many times to maximize their yield, making the mapping step iterative. As a result, robust, efficient, yet easy to use bioinformatic pipelines are needed to map and remap the sequencing reads in a reproducible way, while allowing a wide range of users to modify predefined options and navigate through the results to, e.g., select promising samples, extracts, or libraries.

Two mapping pipelines optimized for aDNA have been published: PALEOMIX (Schubert, et al., 2014) which was designed for stand-alone servers (e.g., without a queuing system) and nf-core/eager (Yates, et al., 2021) which is based on the workflow manager Nextflow (Di Tommaso, et al., 2017) and is designed to make use of distributed computer infrastructures used nowadays (e.g., cluster or cloud services). Both tools have common and specific functionalities. For instance, only PALEOMIX allows for mapping to multiple reference genomes and both tools require considerable amounts of storage for temporary files.

Here, we introduce mapache, a pipeline designed to map in a reproducible manner both ancient and present-day DNA sequences to one or multiple reference genomes and to analyse the resulting alignments by outputting summary statistics. The focus of the pipeline is to have a lightweight, efficient, flexible, and easy to use mapping framework. Mapache is implemented in the workflow manager Snakemake (Mölder, et al., 2021) and it thus inherits its flexibility, scalability, and reproducibility. Snakemake is based on Python allowing for experienced users to easily adapt the code or to combine the workflow with other Snakmake workflows. Compared to PALEOMIX, mapache can run on a single workstation, a HPC system or in the cloud, and compared to nf-core/eager, mapache can map reads to multiple reference genomes. Moreover, unlike the other tools, mapache manages temporal files efficiently, keeping the storage space small. Finally, mapache allows imputing (low-coverage) genomes with GLIMPSE (Rubinacci, et al., 2021).

**Reproducibility**
**Scalability.** Mapache scales well for different (but common) research cases: screening for aDNA quality (mapping of many small FASTQ files), mapping of high-coverage genomes (many large FASTQ files) to a single reference, as well as reconstructing microbial genomes from a metagenomic sample (mapping many FASTQ files to many reference genomes). Another important feature in mapache is its good handling of space for intermediate files. By default, these files are automatically erased when no longer needed, which is especially important when storage space is limited. Finally, depending on the input size, mapache can be launched locally (for small datasets) or on a distributed system (preferred setup for medium and large projects). These features make mapache an ideal tool for running large projects on a cluster: the mapping of six high-coverage ancient human genomes (246 GB, 743 million reads; Clemente et al. (2021)) took 2.5 hours and the mapping of 150 modern human samples (15 TB of input data; European Genome-Phenome Archive: EGAD00001007082) took less than a week on a cluster.
**Automation.** Mapache is launched with a single command line and can be re-launched with the same command if for any reason a mapping run fails (e.g., due to a lack of memory or time).
**Portability.** Mapache can easily be installed on any Linux computer. Dependencies are automatically installed using Conda.

**Workflow**
**Configuration file.** Mapache can be customized using a configuration file. Most parameters contain default values optimized for mapping ancient DNA reads to the human reference genome. The configuration is highly flexible, allowing the user to adapt mapache to the input data: 1) some steps of the workflow can be skipped, 2) different software are available for some steps, and 3) any tool-specific parameter may be included in the pipeline. For example, the mapping pipeline can be customized with parameters suited to map present-day DNA; the user can select among BWA-aln, BWA-mem and bowtie2 for mapping (see Table 1 for a complete list of tools and appropriate references); any valid parameter can be passed to AdapterRemoval with the option "adapterremoval_params". Finally, the workflow can be adapted to the available computer infrastructure by specifying the number of threads, amount of memory and running time for each step in the configuration file.

**Default mapping workflow**. A detailed description of all the steps can be found on the wiki and the list of used software is listed in Table 1. The default workflow (Figure 1) is optimized for mapping ancient DNA reads to the human reference genome. In brief, adapters, short reads and low-quality bases are first removed. Reads are mapped with BWA-aln, disabling the seed (-l 1024, Schubert et al. (2012)). The mapped reads are then merged by library and duplicates are removed. The reads are merged by sample and reads are realigned locally around indels. Finally, the MD flag is recomputed. As an option, unmapped/low-quality reads can be stored in a separate BAM file per sample.

**Imputation (optional).** Mapache allows for imputation and phasing of (low-coverage) genomes using GLIMPSE, a tool that is shown to accurately impute both low-coverage present-day (Rubinacci, et al., 2021) and ancient human genomes (Sousa Da Mota, et al., 2022).

**Output.** Mapache's main outputs are the alignments per sample and genome as BAM files and, if requested, a) low-quality and unmapped reads and b) imputed genotypes in VCF format. Mapache outputs summary statistics on the final BAM and intermediate files, including depth of coverage for all or a subset of the chromosomes, genomic sex, damage patterns, and read lengths. Statistics are returned as tables and graphs, allowing to quickly investigate the run. For in-depth analyses, mapache can generate several individual reports (fastqc, qualimap, multiqc). At the end of a run a Snakemake report can be generated which combines all the summary statistics and graphics, runtime statistics (runtime, time of execution, and executed code chunks) in a single self-contained html, which can be easily shared and stored.

**Conclusion**
Mapache is a flexible mapping workflow optimized for runtime and minimal storage usage to map efficiently both ancient and present-day DNA sequences in a reproducible manner. Mapache comes with an extensive wiki and many pre-defined parameters making it easy to get started. Furthermore, as the underlying language is Python, if users want to go beyond the implemented capabilities, it is easy to modify the code or to combine the workflow with any other Snakemake workflow. Finally, mapache includes downstream processes that are also standalone and can be run on existing BAM files.

**Funding and Acknowledgements**
This work has been supported by the Swiss National Science Foundation and a European Research Council grant to ASM. We would like to thank F Michaud and YO Arizmendi Cárdenas for helpful discussions and feedback.

**References**
Andrews, S. FASTQC. A quality control tool for high throughput sequence data. In.; 2010.
Clemente, F., *et al.* The genomic history of the Aegean palatial civilizations. *Cell* 2021;184(10):2565-2586 e2521.

Danecek, P., *et al.* Twelve years of SAMtools and BCFtools. *GigaScience* 2021;10(2).
DePristo, M.A., *et al.* A framework for variation discovery and genotyping using next-generation DNA sequencing data. *Nature Genetics* 2011;43(5):491-+.
Di Tommaso, P., *et al.* Nextflow enables reproducible computational workflows. *Nat Biotechnol* 2017;35(4):316-319.
Ewels, P., *et al.* MultiQC: summarize analysis results for multiple tools and samples in a single report. *Bioinformatics* 2016;32(19):3047-3048.
Jonsson, H., *et al.* mapDamage2.0: fast approximate Bayesian estimates of ancient DNA damage parameters. *Bioinformatics* 2013;29(13):1682-1684.
Langmead, B. and Salzberg, S.L. Fast gapped-read alignment with Bowtie 2. *Nat Methods* 2012;9(4):357-359.
Li, H. and Durbin, R. Fast and accurate long-read alignment with Burrows-Wheeler transform. *Bioinformatics* 2010;26(5):589-595.
Malaspinas, A.S., *et al.* bammds: a tool for assessing the ancestry of low-depth whole-genome data using multidimensional scaling (MDS). *Bioinformatics* 2014;30(20):2962-2964.
Mölder, F., *et al.* Sustainable data analysis with Snakemake [version 1; peer review: awaiting peer review]. *F1000Research* 2021;10(33).
Okonechnikov, K., Conesa, A. and Garcia-Alcalde, F. Qualimap 2: advanced multi-sample quality control for high-throughput sequencing data. *Bioinformatics* 2016;32(2):292-294.
Peltzer, A., *et al.* EAGER: efficient ancient genome reconstruction. *Genome Biology* 2016;17.
Quinlan, A.R. and Hall, I.M. BEDTools: a flexible suite of utilities for comparing genomic features. *Bioinformatics* 2010;26(6):841-842.
R Core Team. 2022. R: A Language and Environment for Statistical Computing. http://www.R-project.org
Rubinacci, S., *et al.* Efficient phasing and imputation of low-coverage sequencing data using large reference panels. *Nature Genetics* 2021;53(1):120-126.
Schubert, M., *et al.* Characterization of ancient and modern genomes by SNP detection and phylogenomic and metagenomic analysis using PALEOMIX. *Nat Protoc* 2014;9(5):1056-1082.
Schubert, M., *et al.* Improving ancient DNA read mapping against modern reference genomes. *BMC Genomics* 2012;13:178.
Schubert, M., Lindgreen, S. and Orlando, L. AdapterRemoval v2: rapid adapter trimming, identification, and read merging. *BMC Res Notes* 2016;9:88.
Sousa Da Mota, B., *et al.* Imputation of ancient genomes. In.: Cold Spring Harbor Laboratory; 2022.
Yates, J.A.F., *et al.* Reproducible, portable, and efficient ancient genome reconstruction with nf-core/eager. *PeerJ* 2021;9.

**Figure 1.** Main workflow of mapache. Each box shows a step and related software. Dashed boxes are optional. The grey box shows standalone steps which can be computed on any BAM file.

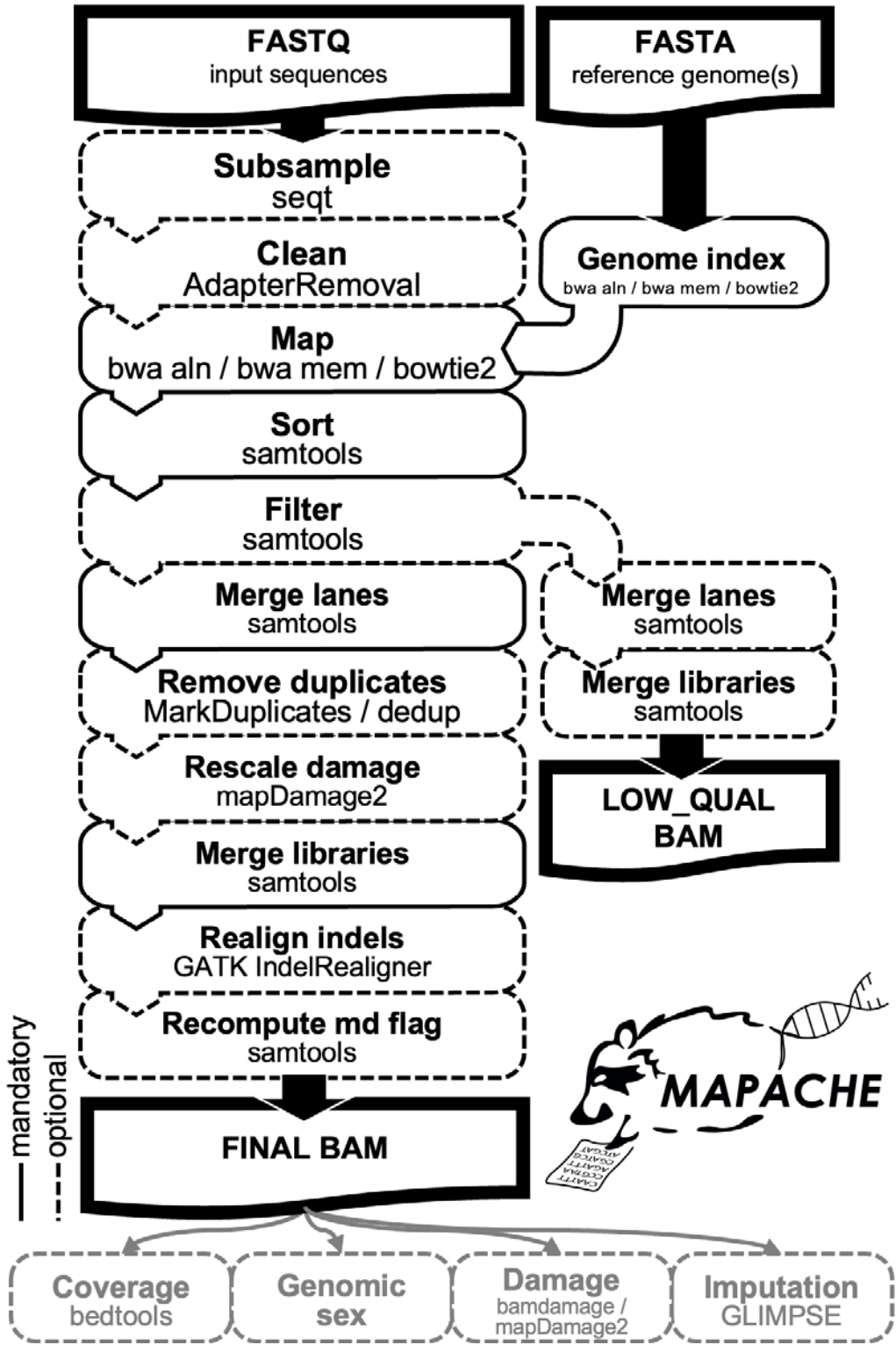

**Table 1.** List of software used for the different steps of the workflow.

| Step | Software | Version | Publication | Link |
|---|---|---|---|---|
|  | Snakemake | 6.4.1 | Mölder, et al. (2021) | https://github.com/snakemake/snakemake |
| Subsample | seqtk | 1.3 |  | https://github.com/lh3/seqtk |
| Clean | AdapterRemoval2 | 2.3.2 | Schubert, et al. (2016) | https://github.com/MikkelSchubert/adapterremoval |
| Map | bwa aln | 0.7.17 | Li and Durbin (2010) | https://github.com/lh3/bwa |
| Map | bwa mem | 0.7.17 | Li and Durbin (2010) | https://github.com/lh3/bwa |
| Map | bowtie2 | 2.4.4 | Langmead and Salzberg (2012) | https://github.com/BenLangmead/bowtie2 |
| Sort | samtools | 1.14 | Danecek, et al. (2021) | https://github.com/samtools/samtools |
| Filter | samtools | 1.14 | Danecek, et al. (2021) | https://github.com/samtools/samtools |
| Merge lanes | samtools | 1.14 | Danecek, et al. (2021) | https://github.com/samtools/samtools |
| Remove duplicates | Picard MarkDuplicates | 2.25.5 | Broad Institute | http://broadinstitute.github.io/picard |
| Remove duplicates | dedup | 0.12.8 | Peltzer, et al. (2016) | https://github.com/apeltzer/DeDup |
| Rescale damage | MapDamage2 | 2.2.1 | Jonsson, et al. (2013) | https://github.com/ginolhac/mapDamage |
| Merge libraries | samtools | 1.14 | Danecek, et al. (2021) | https://github.com/samtools/samtools |
| Realign indels | GATK IndelRealigner | 3.8 | DePristo, et al. (2011) | https://gatk.broadinstitute.org |
| Recompute md flag | samtools | 1.14 | Danecek, et al. (2021) | https://github.com/samtools/samtools |
| Imputation | GLIMPSE | 1.1.1 | Rubinacci, et al. (2021) | https://github.com/odelaneau/GLIMPSE |
| Imputation | bcftools | 1.15 | Danecek, et al. (2021) | https://github.com/samtools/bcftools |
| Reports | fastqc | 0.11.9 | Andrews (2010) | https://www.bioinformatics.babraham.ac.uk/projects/fastqc |
| Reports | qualimap | 2.2.2d | Okonechnikov, et al. (2016) | http://qualimap.conesalab.org |
| Reports | multiQC | 1.11 | Ewels, et al. (2016) | https://multiqc.info |
| Statistics | bedtools | 2.30.0 | Quinlan and Hall (2010) | https://github.com/arq5x/bedtools2 |
| Statistics | bamdamage | modified | Malaspinas, et al. (2014) | https://savannah.nongnu.org/projects/bammds |
| Statistics | R | 4.0 | R Core Team (2022) | https://www.r-project.org |